\documentclass[reprint,amsmath,amssymb,aps,onecolumn,floatfix]{revtex4}

\usepackage{graphicx}
\usepackage{xcolor}
\usepackage{amsfonts}
\usepackage{amssymb} 
\usepackage{amsmath}
\usepackage{hyperref}
\usepackage{natbib}
\usepackage{float}
\usepackage{dcolumn}
\usepackage{bm}
\usepackage{latexsym,color}

\DeclareUnicodeCharacter{2212}{-}

\newcommand{\old}[1]{}

\def\be{\begin{equation}}
	\def\ee{\end{equation}}
\def\bea{\begin{eqnarray}}
	\def\eea{\end{eqnarray}}

\begin{document}

\preprint{APS/123-QED}

\title{Circular geodesics in a New Generalization of q-metric}

\author{Shokoufe Faraji}
\email{shokoufe.faraji@zarm.uni-bremen.de}
\affiliation{%
University of Bremen, Center of Applied Space Technology and Microgravity (ZARM), 28359 Germany}%

\begin{abstract}
This paper introduces an alternative generalization of the static solution with quadrupole moment, the q-metric, that describes a deformed compact object in the presence of the external fields characterized by multipole moments. In addition, we also examine the impact of the external fields up to quadrupole on the circular geodesics and the interplay of these two quadrupoles on the place of innermost stable circular orbit (ISCO) in the equatorial plane.
\end{abstract}
\maketitle

\section{Introduction} 
In general relativity the exact and approximate solutions describing a real source is always of high interest. Here our focus is on solutions considering quadrupole moment to describe vacuum static axisymmetric solutions with mass and different quadrupole parameters.
In this respect, the first static and axially symmetric solutions with arbitrary quadrupole moment are described by Weyl in 1917 \cite{doi:10.1002/andp.19173591804}. Then Erez and Rosen discovered static solutions with arbitrary quadrupole in prolate spheroidal coordinates in 1959 \cite{osti_4201189}. Later Zipoy \cite{doi:10.1063/1.1705005}, and Voorhees \cite{PhysRevD.2.2119} found a transformation that generates the simplest solution which is known as $\gamma$-metric, and later on as $\rm q$-metric. This metric possesses interesting physical aspects and in the case of exact spherical symmetry it reduces to the Schwarzschild metric. 

The above procedure, however, leads to having Minkowski space as the limiting case. While this seems natural in the first place, the question of how an external distribution of mass may distort compact objects might be of some interest. In this perspective, in this paper, the static $\rm q$-metric has been generalized by considering an external matter distribution up to quadrupoles, similar to adding a magnetic environment to the black hole solution \cite{1976JMP....17...54E}. Solutions of the static Einstein vacuum equations obtained by Weyl's method for a long time have played a relatively important role in the describing exterior gravitational field of axially symmetric compact objects. In the present work, utilizing the Weyl's method we used the $\rm q$-metric as a background metric to present a family of solutions, given as an expansion in terms of Legendre polynomials, that may describe for instance the exterior gravitational field of a non-spherically symmetric body embedded in a external gravitational field . This generalization preserves the main virtues of the seed metric aside from asymptotically flatness by its construction.

Regarding, this procedure of considering gravitational field of surrounding matter via quadrupole was pioneered by
the work of Doroshkevich and his colleagues in 1965 considering an external gravitational field up to a quadrupole to the Schwarzschild space-time \cite{1965ZhETF...49.170D}. However, a detailed analysis of the distorted Schwarzschild space-time's global properties was introduced in 1982 by Geroch \& Hartle \cite{1982JMP....23..680G}. Later the explicit form of this metric generalized to Kerr black hole  \cite{1997PhLA..230....7B}.

Thus, it would be of apparent interest to derive the explicit set of metric functions describing a family of $\rm q$-metric in a static external gravitational field where considered in the present work. The first reason for choosing $\rm q$-metric to work with is its mathematical structure that facilitates its study, and let to derive and analysis the equations. Furthermore, for instance, in the relativistic astrophysical study, it is assumed that astrophysical compact objects are described by the Schwarzschild or Kerr space-times. However, besides these relevant solutions, others can imitate a black hole's properties e.g. \cite{PhysRevD.78.024040,10.1093/mnras/stz219}.

Moreover, in the astrophysics area, people attempt to determine the observable predictions of strong-field images of accretion flows in many ways. In this respect, this approach may provide an opportunity to take quadrupole moments as the additional physical degrees of freedom not only to the central compact object but its surroundings. For example, testing quasi-periodic oscillations in this background, shows its ability to connect the data to the model which is a work in progress. In addition, the properties of the thin accretion disc model located in this space-time depends explicitly on the value of both quadrupole parameters in a way that it is always possible to distinguish between a distorted Schwarzschild black hole and a distorted, deformed compact object, which is the work in progress. Besides, there is no doubt on the fundamental importance of gravitational waves in physics, where the experimental evidence finally supported the purely theoretical research in this area \cite{PhysRevLett.116.061102}. 
As discussed in \cite{1995PhRvD..52.5707R}, in the case of extreme mass ratio inspiral (EMRI), one can extract the multipole moments from the gravitational wave signal, and any non-Kerr multipole moments should be encoded in the waves. Therefore, we can expect that this metric may be applicable in the study of gravitational waves generated in an EMRI.  

In the second part of the present work, we consider this metric up to quadrupole to  carry out the characterization of the impact of both quadrupole parameters on the circular geodesics in the equatorial plane. Of course, the effect of other multipole moments are negligible comparing to quadrupole moments. Indeed, the properties of congruences of circular and quasi-circular orbits in an axisymmetric background seems vital to comprehend accretion processes in the vicinity of compact objects. The stable circular orbits, in particular is important to study the accretion discs in the vicinity of a compact object. They encode information about the possibility of existence of the discs. Further, the inner boundary of the thin accretion disc is determined by an innermost stable circular orbit with a high accuracy \cite{2003ApJ...592..354A}. Moreover, the radial and vertical epicyclic frequencies are the most important characteristics of these orbits that are crucial to understanding observational phenomena such as quasi-periodic oscillations, which is a quite profound puzzle in the x-ray observational data of accretion discs \cite{2000ARA&A..38..717V,2006csxs.book..157M,INGRAM2019101524}.

As the next step of this work, one can generalized this approach to include  rotation in this setup to describe more realistic scenarios. However, still it is possible to explain some of the observational data within a static setup. For example, it has been shown that the possible resonant oscillations of the thick accretion disc can be observed even if the source of radiation is steady and perfectly axisymmetric \cite{2004ApJ...617L..45B}.

The paper's organization is as follows: the $\rm q$-metric is briefly explained in Section \ref{qmetric}. The generalized $\rm q$-metric is introduced in Section \ref{general}, and the circular geodesic on the equatorial plane in this background is discussed in Section \ref{ISCOsec}. The summary is presented in Section \ref{discuss}. In this paper, the metric signature convention $(-,+,+,+)$ and the geometrized units where $c=1$ and $G=1$, are used.
\section{\rm q-metric}\label{qmetric}

The Weyl's family of solutions to the static Einstein vacuum equations, have been used to  modeling the exterior gravitational field of compact axially symmetric bodies. In this regard, the $\rm q$-metric describes static, axially symmetric, and asymptotically flat solutions to the Einstein equation with quadrupole moment generalizing the Schwarzschild family. The metric represents the exterior gravitational field of an isolated static axisymmetric mass distribution. It can be used to investigate the exterior fields of slightly deformed astrophysical objects in the strong-field regime \cite{Quevedo:2010vx}. In fact, the presence of a quadrupole, can change the geometric properties of space-time drastically (see e.g. \cite{2021A&A...654A.100F}). The metric in the prolate spheroidal coordinates \footnote{Prolate spheroidal coordinates are three-dimensional orthogonal coordinates that result from rotating the two-dimensional elliptic coordinates about the focal axis of the ellipse.} is presented as follows \cite{osti_4201189,PhysRevD.39.2904}

\begin{align}\label{Imetric}
	d s^2 &=- \left( \frac{x-1}{x+1} \right)^{(1+\alpha)} d t^2+ M^2 (x^2-1)\left( \frac{x+1}{x-1} \right)^{(1+\alpha)}\nonumber\\
	 &\left[\left( \frac{x^2-1}{x^2-y^2}\right)^{\alpha(2+\alpha)} \left( \frac{d x^2}{x^2-1}+ \frac{d y^2}{1-y^2} \right)\right.\nonumber\\
	 &\left.+ (1-y^2) d\phi^2 \right],\
\end{align}
where $t \in (-\infty, +\infty)$, $x \in (1, +\infty)$, $y \in [-1,1]$, and $\phi \in [0, 2\pi)$. Besides, $M$ is a parameter with dimension of length and $m=M(1+\alpha)$ gives the lowest multipole moment by Geroch definition \cite{1970JMP....11.2580G}. Indeed, $m$ be positive is a necessary condition to avoid having a negative mass distribution \footnote{ In fact, the Arnowitt-Deser-Misner mass which characterizes the physical properties of the exact solution also has the same expression and should be positive \cite[Appendix]{PhysRevD.93.024024}.}. It is equivalent to restricting quadrupole at most to $\alpha\in(-1,\infty)$. For vanishing $\alpha$ the Schwarzschild space-time is recovered, also for $\alpha=-1$ all multipole moments vanish and the space-time will be flat. The relation between this coordinates system and the Schwarzschild like coordinates is given by


\begin{align}\label{transf1}
 x =\frac{r}{M}-1 \,, \quad  y= \cos\theta\,,
\end{align}
and the metric reads as  \cite{2011IJMPD..20.1779Q}
\begin{align}
   ds^2 = &\left(1-\frac{2M}{r}\right)^{1+\alpha} dt^2 - \left(1-\frac{2M}{r}\right)^
    {-\alpha} \nonumber\\
    &\left[ \left(1+\frac{M^2\sin^2\theta}{r^2-2Mr}\right)^{-\alpha(2+\alpha)}\left(\frac{dr^2}{1-\frac{2M}{r}}+r^2d\theta^2\right)\right. \nonumber\\
  &\left.+r^2\sin^2\theta d\phi^2\right].\
\end{align}
This metric has a central curvature singularity at $x=-1$ (or $r=0$). Moreover, an additional singularity appears at $x=1$ (or $r=2M$), and the norm of the time-like Killing vector at this radius vanishes. However, outside this hypersurface, there exists no additional horizon. Nevertheless, considering a relatively small quadrupole moment, a physically interior solution can cover this hypersurface, since it is closely place to the central singularity \cite{Quevedo:2010vx} . Besides, out of this region, there is no more singularity, and the metric is asymptotically flat. 


\section{Generalized q-metric}\label{general}
In this section, we start with the static and axisymmetric solutions which are described via the Weyl metric

\begin{align}\label{weylmetric}
d s^2 & =-e^{2\psi}d t^2 +e^{2(\gamma-\psi)}(d\rho^2+d z^2) +e^{-2\psi}\rho^2 d\phi^2\,,
\end{align}
where $\psi=\psi(\rho,z)$ and $\gamma=\gamma(\rho,z)$ are the metric functions, and $\psi$ plays the role of the gravitational potential. 
If we consider a three-dimensional manifold $N$, orthogonal to the static Killing vector field, then this metric induces the flat metric on $N$. Besides, the metric function $\psi$ with respect to this flat metric obeys the Laplace equation



\begin{align}\label{laplap}
\psi_{,\rho\rho}+\frac{1}{\rho}\psi_{,\rho}+\psi_{,zz}=0,
\end{align}
where $","$ is used for partial derivatives. This linear equation is the key factor in the Weyl technique of generating solutions. The metric function $\gamma$ is obtained by the explicit form of the function $\psi$, and the equation \eqref{laplap} is the integrability condition for this solution


\begin{align} \label{equgamma}
  \nabla \gamma \nabla \rho = \rho (\nabla \psi)^2,
\end{align}
where $\nabla :=\partial_z+i\partial_{\rho}$, or equivalently

\begin{align}\label{gammaeq}
\gamma_{,\rho}=&\rho(\psi^2_{,\rho}-\psi^2_{,z}),\nonumber\\
\gamma_{,z}=&2\rho_{}\psi_{,\rho}\psi_{,z}.\
\end{align}
Before proceeding with the $\rm q$-metric, we write field equations in the more symmetric form in the prolate spheroidal coordinates $(t, x, y, \phi)$.  The relation of the cylindrical coordinates of Weyl to the prolate spheroidal coordinates is given by

 \begin{align}
    {x} & =\frac{1}{2{M}}(\sqrt{\rho^2+({z}+{M})^2}+\sqrt{\rho^2+({z}-{M})^2})\nonumber\,,\\
    {y} & =\frac{1}{2{M}}(\sqrt{\rho^2+({z}+{M})^2}-\sqrt{\rho^2+({z}-{M})^2}).
\end{align}
In this coordinates system the field equation for $\psi$ \eqref{laplap} reads as

\begin{align} \label{laplapsym}
    \left[(x^2-1)\psi_{,x}\right]_{,x} + \left[(1-y^2)\psi_{,y}\right]_{,y} =0,
\end{align}
And for $\gamma$ \eqref{gammaeq} is given by

\begin{align} \label{symgamma}
    \gamma_{,x}&=\frac{1-y^2}{x^2-y^2}\left[x(x^2-1)\psi^2_{,x}-x(1-y^2)\psi^2_{,y}-2y(x^2-1)\psi_{,x}\psi_{,y}\right],\nonumber\\
    \gamma_{,y}&=\frac{1-y^2}{x^2-y^2}\left[y(x^2-1)\psi^2_{,x}-y(1-y^2)\psi^2_{,y}+2x(1-y^2)\psi_{,x}\psi_{,y}\right].\
\end{align}
We can see from the form of equation \eqref{laplapsym}, that it allows separable solutions. Therefore, $\psi$ can be written in terms of multiplication of two functions, say $\psi=\mathcal{X}(x)\mathcal{Y}(y)$. By using the separation of variable methods in differential equation, it is easy to see that one can write the $y$ part in terms of Legendre polynomial. Then, the $x$ dependence part, also fulfills the Legender's equation  \cite{Abramowitz:1974:HMF:1098650}

\begin{align}
    \left[(x^2-1)\mathcal{X}_{,x}\right]_{,x}-n(n+1)\mathcal{X}=0.
\end{align}
In brief, $\mathcal{X}$ can be expressed in terms of Legendre polynomials of the first kind and second kind. However, by relaxing the assumption of asymptotic flatness, the second kind’s coefficient should vanish. In addition, by the requirement of elementary flatness \footnote{ In general, these fields should be regular at the symmetry axis. Sometimes this condition is referred to as the elementary flatness condition.} in the neighborhood of the symmetry axis, the general solution for $\psi$ is obtained as

\begin{align}\label{say}
   \psi = \sum_{n=0} \beta_n R^n P_n(\frac{xy}{R}),  \quad R=\sqrt{x^2+y^2-1}\,,
\end{align}
where $P_n$ is Legendre polynomial of order $n$, and $\beta_n\in \mathbb{R}$. Note that, since $\psi$ is the solution of linear Laplace equation any superposition of solutions is still a solution of this equation. Regarding our static metric should represent the $\rm q$-metric in the presence of the external field, so we shall choose field in this form

\begin{align} \label{finalpsi}
    \psi= \frac{(1+\alpha)}{2}\ln\left(\frac{x-1}{x+1}\right) +  \sum_{n=1} \beta_n R^n P_n(\frac{xy}{R}).
\end{align}
To obtain the field function $\gamma$ corresponding to this potential $\psi$ \eqref{finalpsi} explicitly, one needs to solve equations \eqref{symgamma}, where determines $\gamma$ up to some constant. However, the requirement of elementary flatness in the neighborhood of the symmetry axis fixes the constant, and it should be set equal to zero. For the resulting function we obtain this expression

\begin{align}\label{finalgamma}
\gamma & = \frac{(1+\alpha)^2}{2}\ln\left(\frac{x^2-1}{x^2-y^2}\right)+\sum_{n=1} \beta_n (1+\alpha) \sum_{l=0}^{n-1} \left[ (-1)^{n-l+1} (x+y)-x+y \right] R^l P_l \nonumber \\
& \quad + \sum_{k,n=1} \frac{n k \beta_n \beta_k}{(n+k)} R^{n+k} [P_n P_k -P_{n-1} P_{k-1} ]\,, 
\end{align}
where
\begin{align}
P_n:= P_n(\frac{xy}{R})\,, \quad R=\sqrt{x^2+y^2-1}\,.
\end{align} 
Where the first term in \eqref{finalgamma} is the $\gamma$ function of the $\rm q$-metric. For simplicity and emphasis on the external contributions by noting them as $\hat{\psi}$ and $\hat{\gamma}$, we can show the equations \eqref{finalpsi} and \eqref{finalgamma} by

\begin{align}
    \psi&=\psi_{\rm q} + \hat{\psi},\nonumber\\
    \gamma&= \gamma_{\rm q} + \hat{\gamma},\
\end{align}
where $\psi_{\rm q}$ and $\gamma_{\rm q}$ are fields for $\rm q$-metric; namely, one preserves the $\rm q$-metric fields by taking $\hat{\gamma}=\hat{\psi}=0$, equivalently no external field. It can easily be checked that in the limits $\hat{\psi}=0$, $\hat{\gamma}=0$, and $\alpha=0$ we recover the Schwarzschild fields. Ultimately, the metric is then given by

\begin{align}\label{EImetric}
	d s^2 &= - \left( \frac{x-1}{x+1} \right)^{(1+\alpha)} e^{2\hat{\psi}} d t^2+ M^2(x^2-1) e^{-2\hat{\psi}} \nonumber\\
	 &\left( \frac{x+1}{x-1} \right)^{(1+\alpha)}\left[ \left(\frac{x^2-1}{x^2-y^2}\right)^{\alpha(2+\alpha)}e^{2\hat{\gamma}}\right.\nonumber\\
	 &\left. \left( \frac{d x^2}{x^2-1}+\frac{d y^2}{1-y^2} \right)+(1-y^2) d{\phi}^2\right],\
\end{align}
where $t \in (-\infty, +\infty)$, $x \in (1, +\infty)$, $y \in [-1,1]$, and $\phi \in [0, 2\pi)$. Again for $\hat{\psi}=0$, $\hat{\gamma}=0$ the $\rm q$-metric \eqref{Imetric} is recovered. If also $\alpha=0$, the Schwarzschild metric is obtained. Of course, by replacing $\alpha$ with parameter of ZV space-time via $\delta:=1+\alpha$, one can consider this metric as the generalization of ZV space-time, as well. Up to the quadrupole $\beta:=\beta_2$, the external field terms read as follows

\begin{align} \label{1111}
\hat{\psi} & = -\frac{\beta}{2}\left[-3x^2y^2+x^2+y^2-1\right],\nonumber\\
\hat{\gamma} & = -2x\beta(1+\alpha)(1-y^2)\nonumber\\
  &+\frac{\beta^2}{4}(x^2-1)(1-y^2)(-9x^2y^2+x^2+y^2-1).\
  \end{align}
Therefore, if we consider the external fields up to quadrupole, this metric contains three free parameters: the total mass, the deformation parameter $\alpha$, and the distortion $\beta$, which are taken to be relatively small, and connected to the compact object's deformation and the external mass distribution, respectively. Moreover, further analysis shows that these parameters are not independent of each other, as we will see in the following parts. The result is locally valid by its construction, and can be considered as the distorted $\rm q$-metric. From the physics perspective, this solution is similar to the $\rm q$-metric solution with an additional external gravitational field, like adding a magnetic surrounding \cite{1976JMP....17...54E}.

\subsection{Effective potential} \label{circular}
To elucidate some aspects of the influence of the parameters, we consider the effective potential of geodesic motion in this space-time. Regarding the symmetries in the metric, there are two constants of geodesic motion $E$ and $L$

\begin{align} \label{EL}
E&= -{\rm g}_{t t} \dot{t}=\left( \frac{x-1}{x+1} \right)^{(1+\alpha)}e^{2\hat{\psi}} \dot{t},\\
L&= {\rm g}_{\phi \phi}\dot{\phi}=M^2(x^2-1)(1-y^2) \left( \frac{x+1}{x-1} \right)^{(1+\alpha)}e^{-2\hat{\psi}}\dot{\phi},\nonumber\
\end{align}
where "over-dot" notation is used for partial derivatives with respect to the proper time. By using these relations and the normalization condition ${\rm g}_{\rho \nu}\dot{x}^{\rho}\dot{x}^{\nu}=-\epsilon,$ where $\epsilon$ can take values $-1$, $0$ and $1$, for the space-like, light-like and for the time-like trajectories, respectively; the geodesic equation is obtained as

\begin{align}\label{asli}
\left(\frac{x^2-1}{x^2-y^2}\right)^{\alpha(2+\alpha)}M^2e^{2\hat{\gamma}} \dot{x}^2+V^2=E^2,
\end{align}
where

\begin{align}\label{VtwoEI}
V^2&= \frac{x^2-1}{1-y^2}\left(\frac{x^2-1}{x^2-y^2}\right)^{\alpha(2+\alpha)}M^2e^{2\hat{\gamma}}\dot{y}^2\nonumber\\
&+\left( \frac{x-1}{x+1} \right)^{(2\alpha+1)}\frac{L^2e^{4\hat{\psi}}}{M^2(1-y^2)(x+1)^{2}}\nonumber\\
&+\left( \frac{x-1}{x+1} \right)^{(\alpha+1)}e^{2\hat{\psi}}\epsilon.
\end{align}
One can interpret \eqref{asli} as the motion along the $x$ coordinate in terms of $V^2$ so-called potential. However, due to the appearance of $\dot{y}$ in this expression, in fact it is not a potential. In the next subsection, we rewrite $V^2$ in the equatorial plane $(\dot{y}=0)$. Then, it will have the meaning of related effective potential, and we rename it to $ V_{\rm Eff}$, accordingly. In addition, the Christoffel symbols calculated for this metric are presented in Appendix \ref{circulargeo}.


\section{Circular geodesics in the equatorial plane} \label{ISCOsec}
 In this part we consider the metric functions up to quadrupole to analysing the circular motion and the place of ISCO in the equatorial plane. In the equatorial plane $y=0$, the distortion functions \eqref{1111} up to the quadrupole simplify to
\begin{align}\label{gammapsi1}
	\hat{\psi} & = -\frac \beta{2}({x}^2-1)\,, \nonumber\\
	\hat{\gamma} & = -2\beta (1+\alpha)x+\frac{\beta^2}{4}({x}^2-1)^2\,.
\end{align}
The metric in the equatorial plane is given by
\begin{align}\label{EIeq}
	d s^2 &= - \left( \frac{x-1}{x+1} \right)^{(1+\alpha)} e^{2\hat{\psi}} d t^2+ M^2(x^2-1) e^{-2\hat{\psi}} \nonumber\\
	 &\left( \frac{x+1}{x-1} \right)^{(1+\alpha)}\left[ \left(\frac{x^2-1}{x^2}\right)^{\alpha(2+\alpha)}e^{2\hat{\gamma}}\right.\nonumber\\
	 &\left. \left( \frac{d x^2}{x^2-1}+d y^2 \right)+ d{\phi}^2\right].\
\end{align}
Further, the relation \eqref{asli} is reduced to
\begin{align}
\left(\frac{x^2-1}{x^2}\right)^{\alpha(2+\alpha)}M^2e^{2\hat{\gamma}} \dot{x}^2+V_{\rm Eff}=E^2,
\end{align}
where 

\begin{align}\label{Vei}
    V_{\rm Eff}=& \left( \frac{x-1}{x+1} \right)^{(\alpha+1)} e^{2\hat{\psi}}\nonumber\\
    &\left[ \epsilon+\frac{L^2e^{2\hat{\psi}}}{M^2(x+1)^2} \left( \frac{x-1}{x+1} \right)^\alpha\right].\,
\end{align}
In the rest of this section, circular motion in the equatorial plane up to quadrupole is studied.


\subsection{Circular orbits for the time-like trajectory}\label{ueffsec}

In this type of motion, there is no change in the $x$ direction $\dot{x}=0$, so one can study the motion of test particles in the effective potential \eqref{Vei} equivalently. Where $V_{\rm Eff}$ for $\beta=0$, reduce to the effective potential for $\rm q$-metric, and $\alpha=0$ corresponds to the effective potential for distorted Schwarzschild metric, and in the case of $\alpha=\beta=0$ the effective potential for Schwarzschild space-time.



\begin{figure*}
        \includegraphics[width=0.4\textwidth]{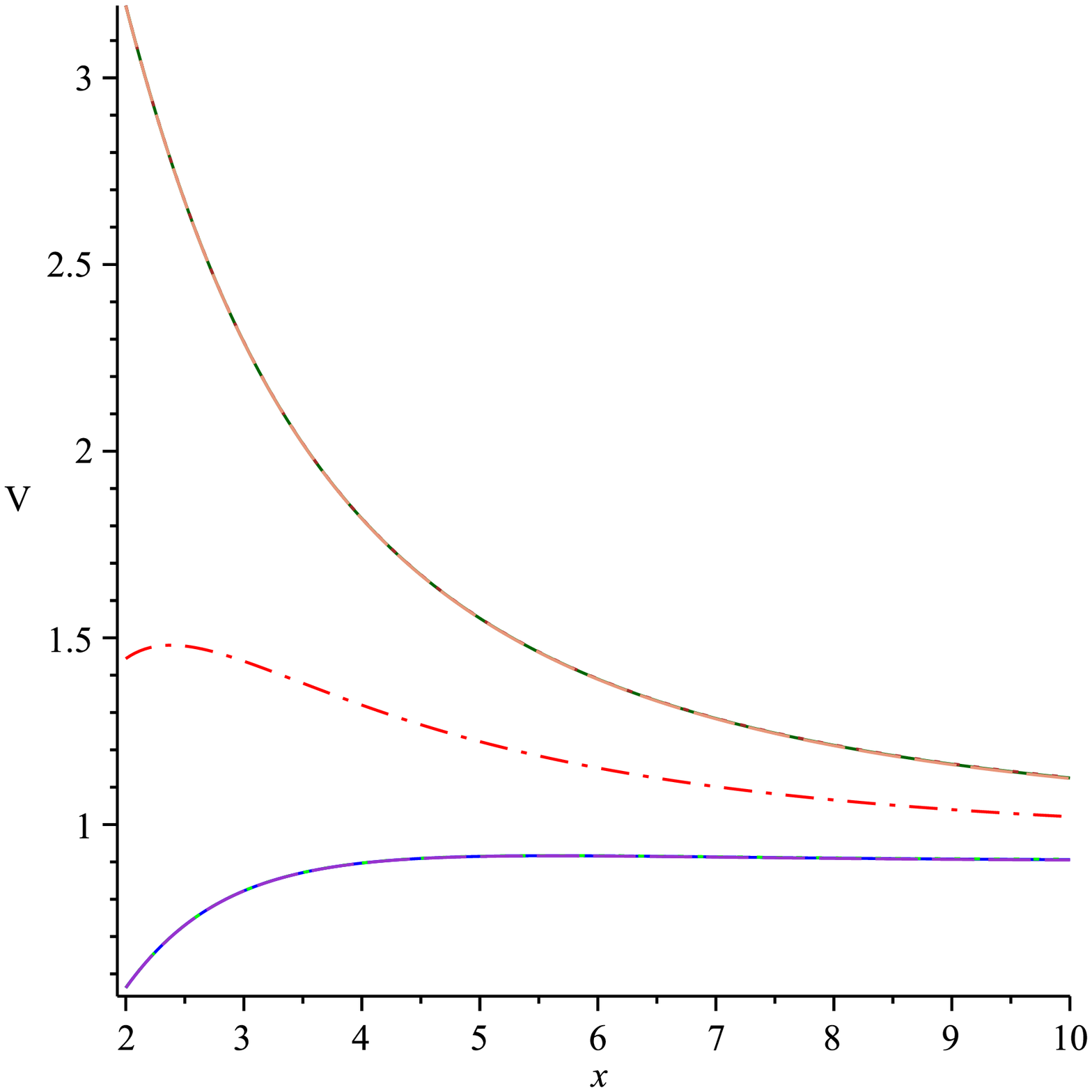}
         \includegraphics[width=0.4\textwidth]{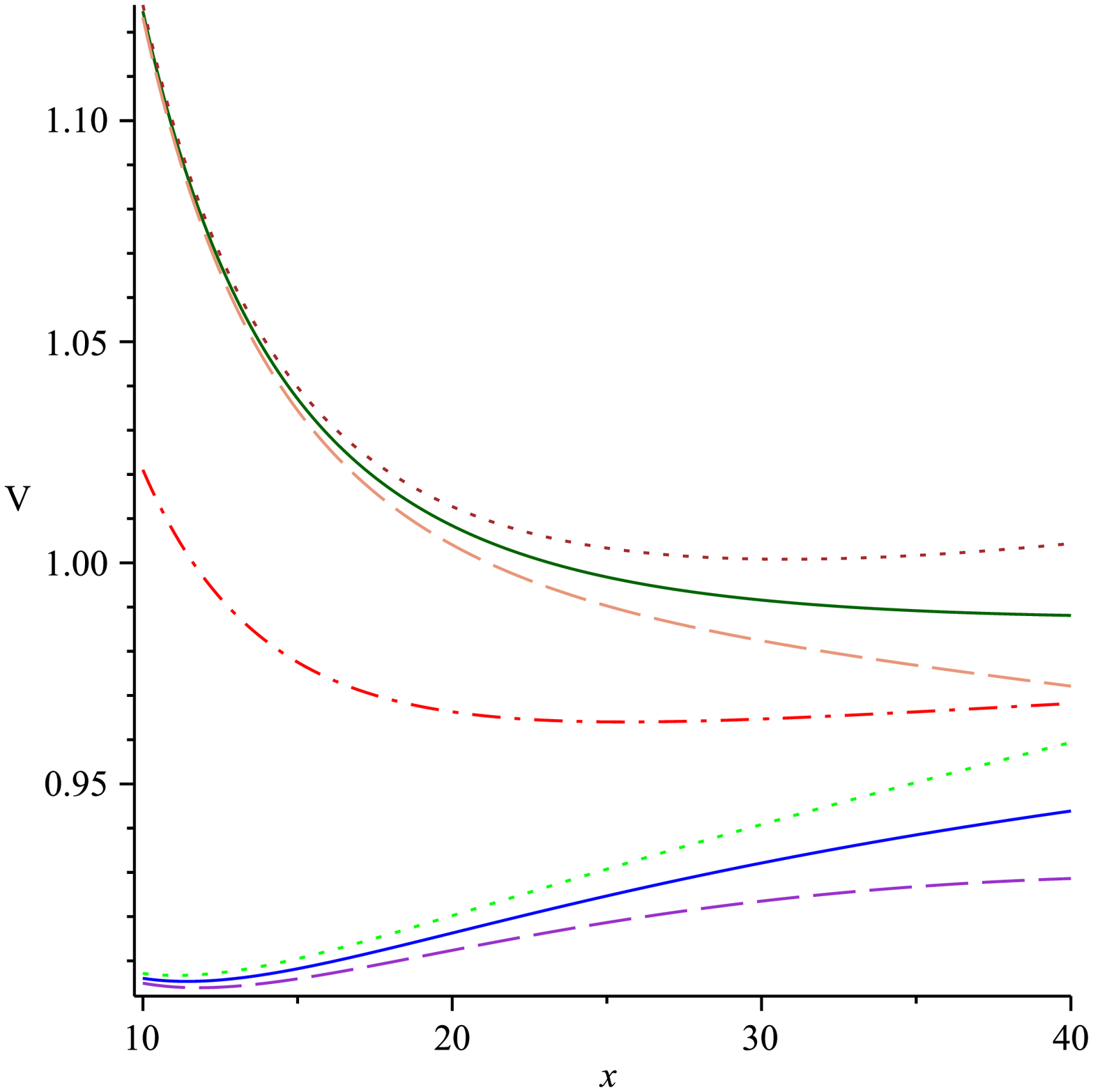}
    \caption{\label{vpqs} Plots of $V_{\rm Eff}$ for different values of $\alpha$ and $\beta$. In both plots, the dot-dashed line in the middle is Schwarzschild $(\alpha=\beta=0)$, the solid line under the Schwarzschild corresponds to values $(\alpha=0.5, \beta=0)$, and the solid line above the Schwarzschild corresponds to $(\alpha=-0.4, \beta=0)$. The dotted line under the Schwarzschild is for the values $(\alpha=0.5, \beta=-0.00001)$, and the dotted line above the Schwarzschild is plotted for $(\alpha=-0.4, \beta=-0.00001)$. The dashed line under the Schwarzschild is for values $(\alpha=0.5, \beta=0.00001)$, and the dashed line above the Schwarzschild is plotted for $(\alpha=-0.4, \beta=0.00001)$.} 
\end{figure*}

To have some examples, in Figure \ref{vpqs}, $V_{\rm Eff}$ is plotted for different values of $\alpha$ and $\beta$. As we see, the Schwarzschild effective potential lies in between the one with a negative value of $\alpha$ and the positive value of $\alpha$. Also, for each fixed value of $\alpha$, the effective potential for a negative $\beta$ is higher than the effective potential for the same $\alpha$ but vanishing $\beta$. The opposite is also true for positive $\beta$ at each $x$. As Figure \ref{vpqs} shows, further away from the central object, they are more diverged.

\begin{figure*}
\centering
\includegraphics[width=0.4\textwidth]{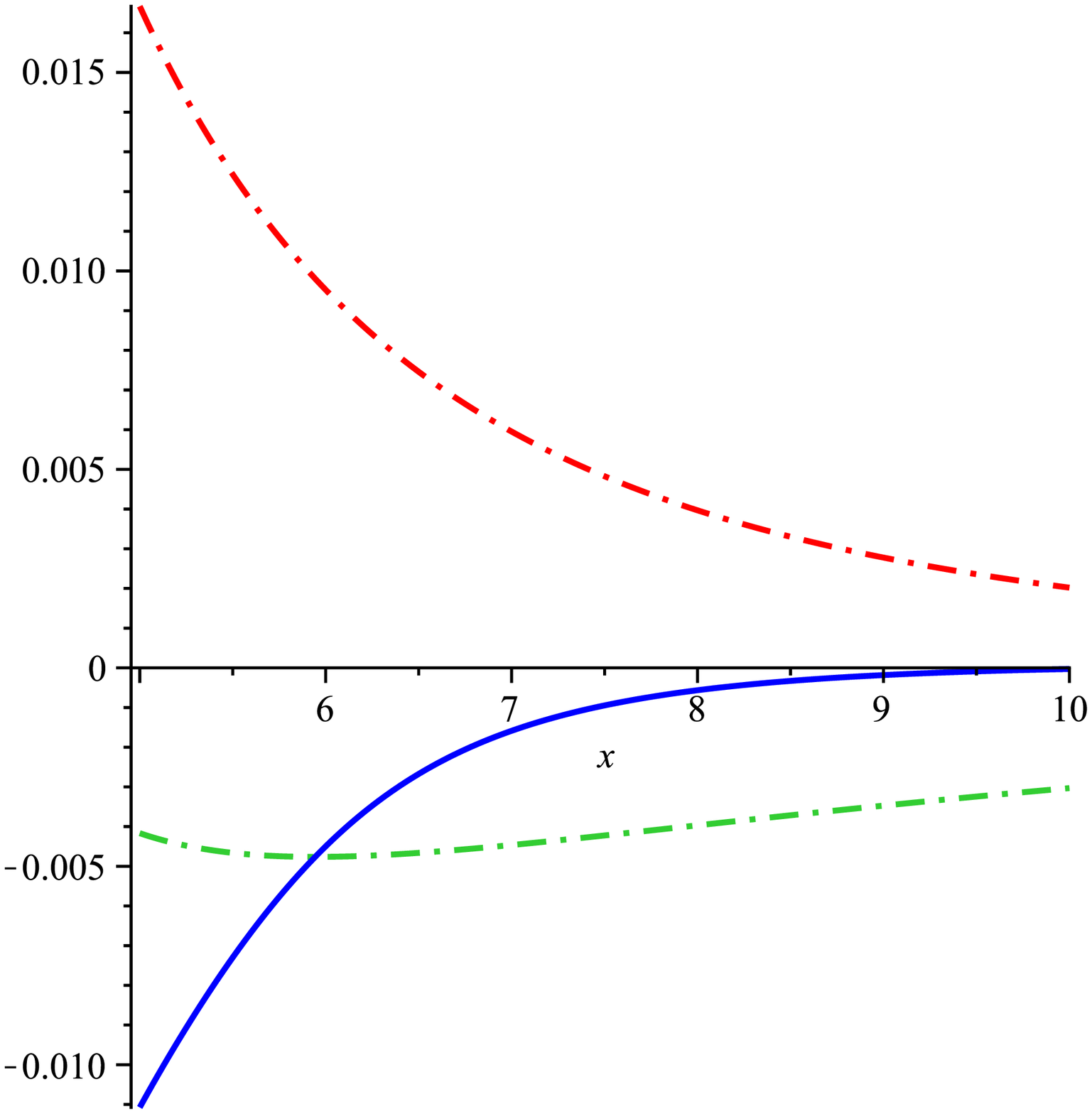}
         \includegraphics[width=0.4\textwidth]{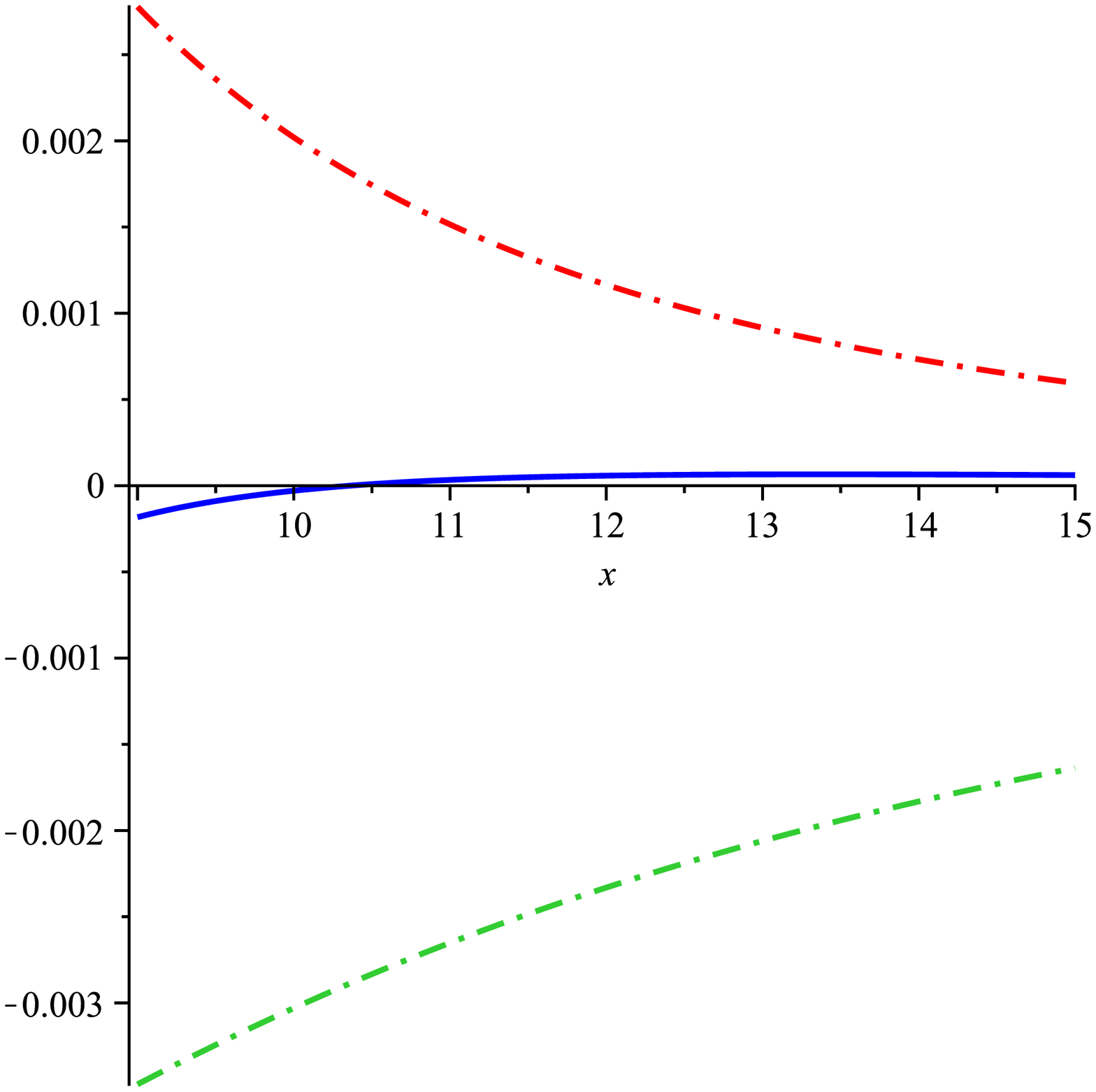}
    \caption{\label{p1q}The dashed lines are $\mathnormal{l}_1$ \eqref{curve1} and $\mathnormal{l}_2$ \eqref{curve2}. The solid lines are the plots of $\beta$ \eqref{qsol} for $\alpha=1$, noted as $\beta^{\alpha=1}$. Minimum of $\beta^{\alpha=1}$ is $-0.0047632$ at $x=5.94338$, and maximum of $\beta^{\alpha=1}$ is $0.0000659$ at $x=13.38972$. Moreover, $\beta=0$ at $x=10.35890$, so this is the place of ISCO for $\alpha=1$ with vanishing $\beta$ in the q-metric.}
\end{figure*}

\begin{figure*}
\centering
        \includegraphics[width=0.4\textwidth]{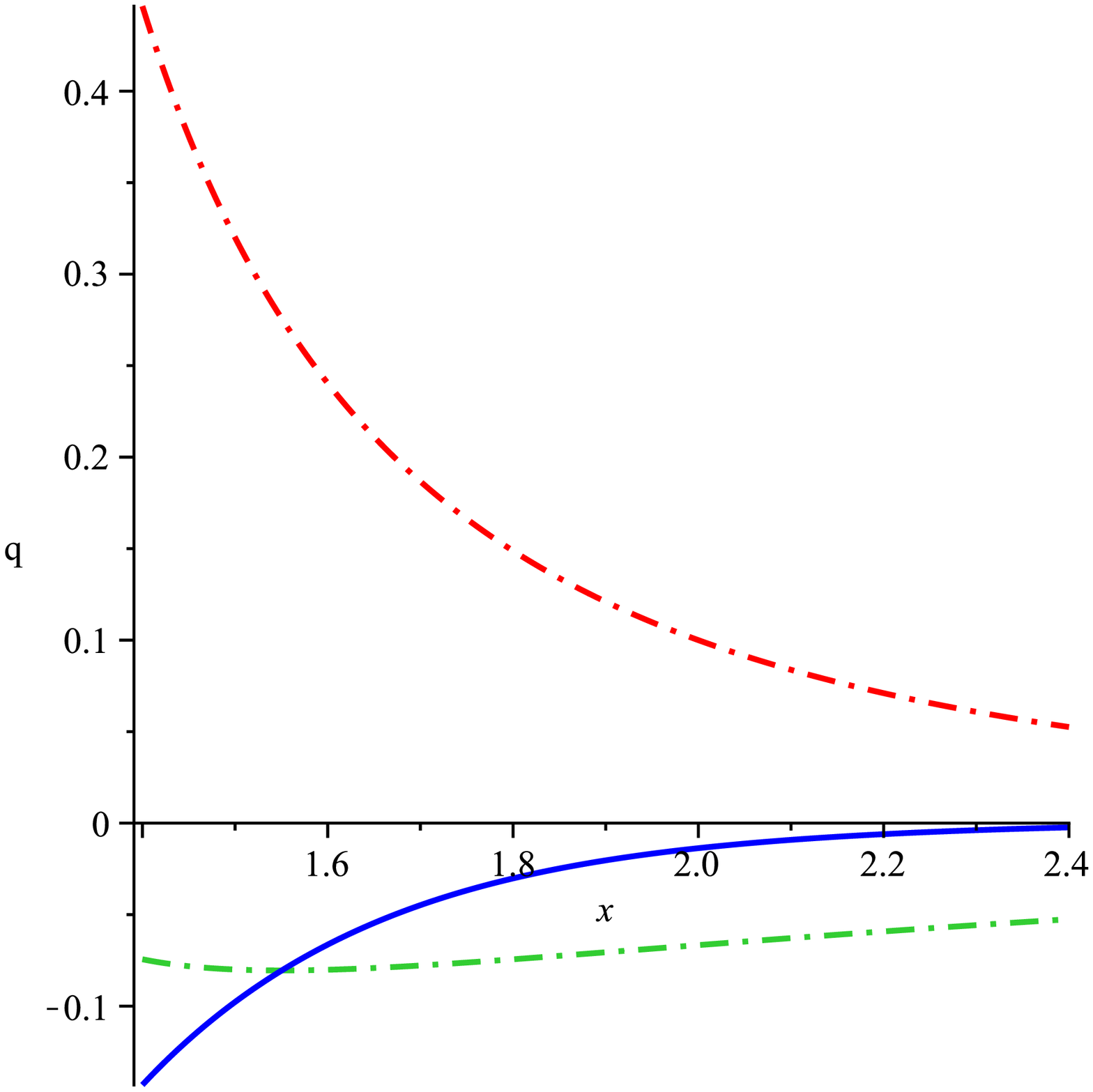}
         \includegraphics[width=0.4\textwidth]{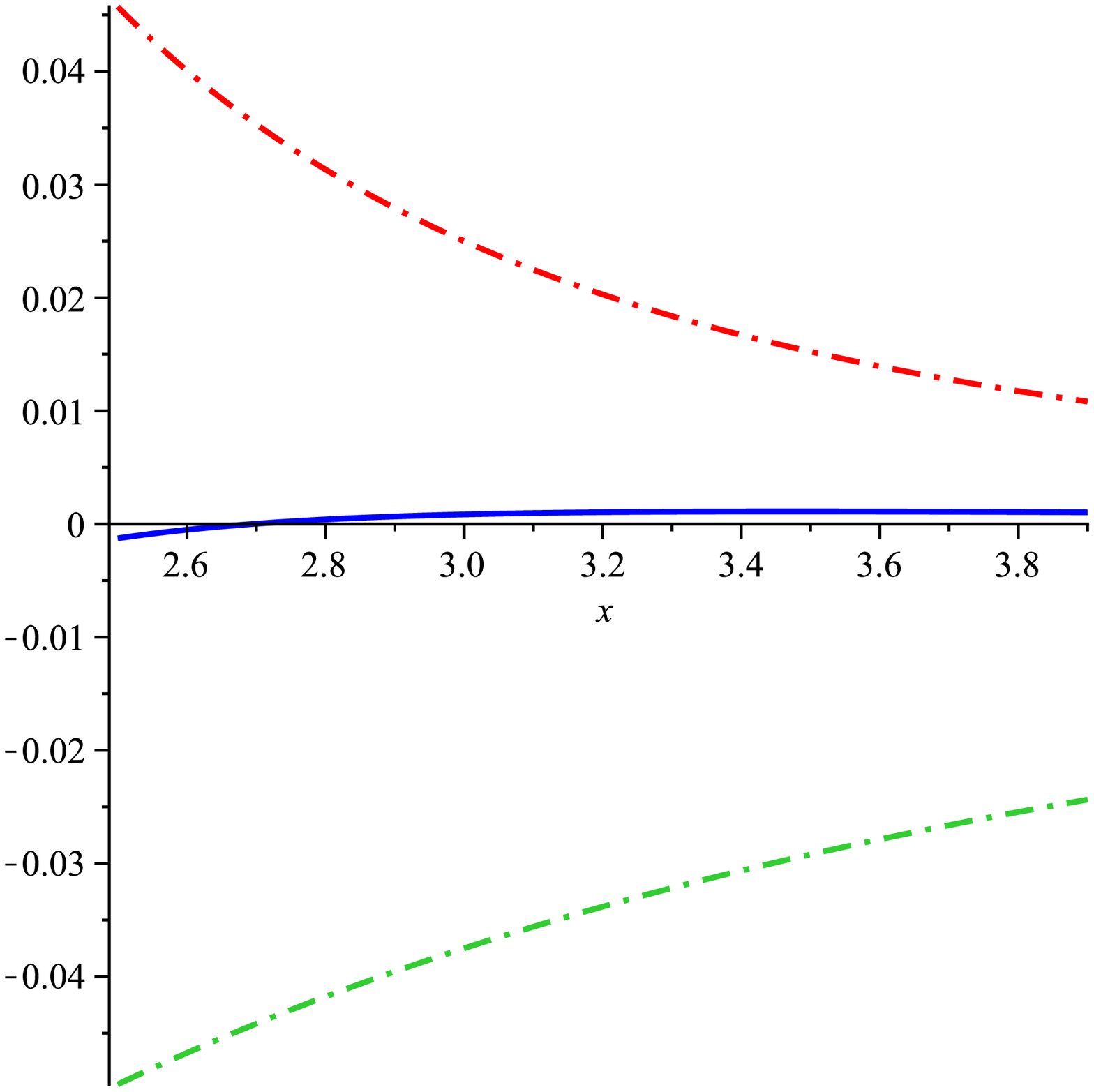}
    \caption{\label{p3q} The dot-dashed lines depict $\mathnormal{l}_1$ \eqref{curve1} and $\mathnormal{l}_2$ \eqref{curve2}. The solid lines present plots of $\beta$ \eqref{qsol} for $\alpha=-0.4$, noted as $\beta^{\alpha=-0.4}$. Minimum of $\beta^{\alpha=-0.4}$ is $-0.0805014$ at $x=1.55038$, and maximum of $\beta^{\alpha=-0.4}$ is $0.0011090$ at $x=3.47165$. Also, $\beta=0$ at $x=2.69443$, so this $x$ shows the value of ISCO for $\alpha=-0.4$ with $\beta=0$ in the q-metric.}
\end{figure*}

\begin{figure}
\centering
        \includegraphics[width=0.3\textwidth]{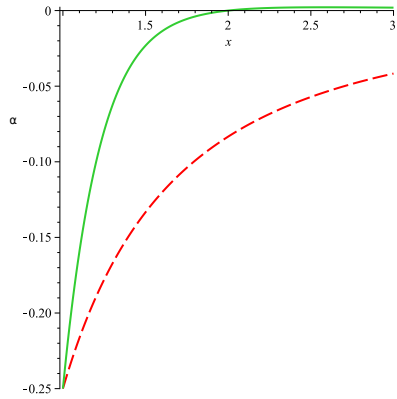}
        \caption{\label{5}Around $\alpha= - 0.5$ curve $\beta$ begins to intersect with curve 1 \eqref{curve1}. Before this value, the minimum is determined by the minimum of $\beta$ itself, and after that, the minimum of $\beta$ is obtained by the intersection of these two curves.}
\end{figure}

Typically, the place of ISCO, the last innermost stable orbit, is the place of the extrema of $V_{\rm Eff}$ and $(V_{\rm Eff})^{'}$ simultaneously. However, finding extrema of $V_{\rm Eff}$, equivalents to analysis the extrema of $L^2$ for massive particles which for this metric is obtained as


\begin{align}
L^2=\frac{M^2(x+1)^{\alpha+2}\left[-\beta x^3+\beta x+\alpha+1\right]}{(x-1)^\alpha\left[2\beta x^3+(1-2\beta)x-2\alpha-2\right]}{e}^{-2 \hat{\psi}}.\label{angmom}
\end{align}
The vertical asymptote of this function for $x>1$ is


\begin{align}\label{curve}
 2\beta x^3+(1-2\beta)x-2\alpha-2=0,
\end{align}
which leads to this relation for $\beta$
\begin{align}\label{curve1}
\mathnormal{l}_{1}:=\frac{-x+2(\alpha+1)}{2x(x^2-1)}.
\end{align}
In the space-time with the quadrupole (in general with multipole moments), there is an interesting possibility that there exists a curve in which $L^2$ may vanish along with it. This means that particles are at rest along this curve with respect to the central object. Of course, this is nor the case in Schwarzschild or in the q-metric space-times. In this way, the external matter manifests its existence by neutralizing the central object's gravitational effect at the region determined by this curve. In this case, this happens along $-\beta x^3+\beta x+\alpha+1=0$, which can be written also as

\begin{align}\label{curve2}
\mathnormal{l}_{2}:=\frac{\alpha+1}{x(x^2-1)}.
\end{align}
However it is worth mentioning that, there is just one position for each chosen value of $\beta$. In general, the region between curves $\mathnormal{l}_1$ \eqref{curve1} and $\mathnormal{l}_2$ \eqref{curve2} defines the valid range for the distortion parameter $\beta$, due to the fact that $L^2$ is a positive function. A straightforward calculation of $\frac{d}{dx}L^2=0$ leads to

\begin{align}\label{fp=0}
&4x^3(x^2-1)^3\beta^3\\
&+6x^2\left(-2(x^2-1)^2\alpha+x^5-2x^4-2x^3+4x^2+x-2\right)\beta^2\nonumber\\
&+4x(x^2-1)(3\alpha^2 + 3(2-x)\alpha + x^2 - 3x + 3)\beta\nonumber\\
&-(\alpha+1)x^2+6(\alpha+1)^2x-4\alpha^3-12\alpha^2-13\alpha-5\nonumber\\
&=0.\nonumber \
\end{align}
Which gives the solution for $\beta$ as

\begin{align}\label{qsol}
\beta = \frac{1}{2x(x^2-1)}(\frac{\sqrt[3]{D^2}-3x^2}{3\sqrt[3]{D}}+2-x+2\alpha),
\end{align}
where

\begin{align}\label{D}
D = 27x^3-81x^2(\alpha+1)+27(\alpha+1)+3\sqrt{\Delta},
\end{align}

and

\begin{align}
\Delta &= 84x^6-486x^5(\alpha+1)+729x^4(\alpha+1)^2\nonumber\\
&+162x^3(\alpha+1)-486x^2(\alpha+1)^2+81(\alpha+1)^2.\
\end{align}
An analyze shows that for any value of $\alpha$ chosen in this domain $[-0.5,\infty)$, the minimum of $\beta$ is obtained at the intersection of the curve $\beta$ \eqref{qsol} with $\mathnormal{l}_1$ \eqref{curve1}. 
Besides, the maximum of the curve $\beta$ in the valid region is determined by the maximum of $\beta$ for this chosen $\alpha$.
For instance, in Figure \ref{p1q}, \ref{p3q} and  \ref {5}, $\beta$ and the valid region for one positive and two negative values $\alpha=1$, $\alpha=-0.4$, and $\alpha=-0.5$ are plotted. In the later, the minimum of $\beta$ is obtained by its intersection with $\mathnormal{l}_1$ \eqref{curve1}, placed at the very close to outer singularity. 


\begin{figure}
\centering
        \includegraphics[width=0.3\textwidth]{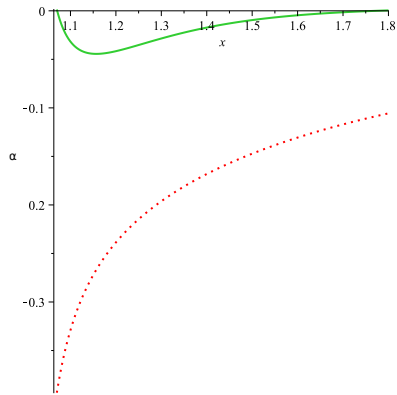}
        \caption{\label{7min} The dotted line is $\mathnormal{l}_1$ \eqref{curve1}, and solid line is $\beta$ \eqref{qsol} for $\alpha=-0.526$. In this interval $[-0.5528,-0.5)$ the minimum is determined by the minimum of the curve $\beta$ \eqref{qsol} itself.}
\end{figure}

For the values of $\alpha$ in this domain $[-1+\frac{\sqrt{5}}{5},-0.5)$, where $-1+\frac{\sqrt{5}}{5}\sim -0.553$, the curve $\beta$ behaves differently, and it always lies in the valid region, so the minimum and maximum of $\beta$ are determined by its extrema. 
In Figure \ref{7min}, the minimum for $\alpha=-0.526$ was shown, which its minimum is obtained by the minimum of a curve $\beta$ itself.


\begin{figure*}
\centering
\includegraphics[width=0.4\textwidth]{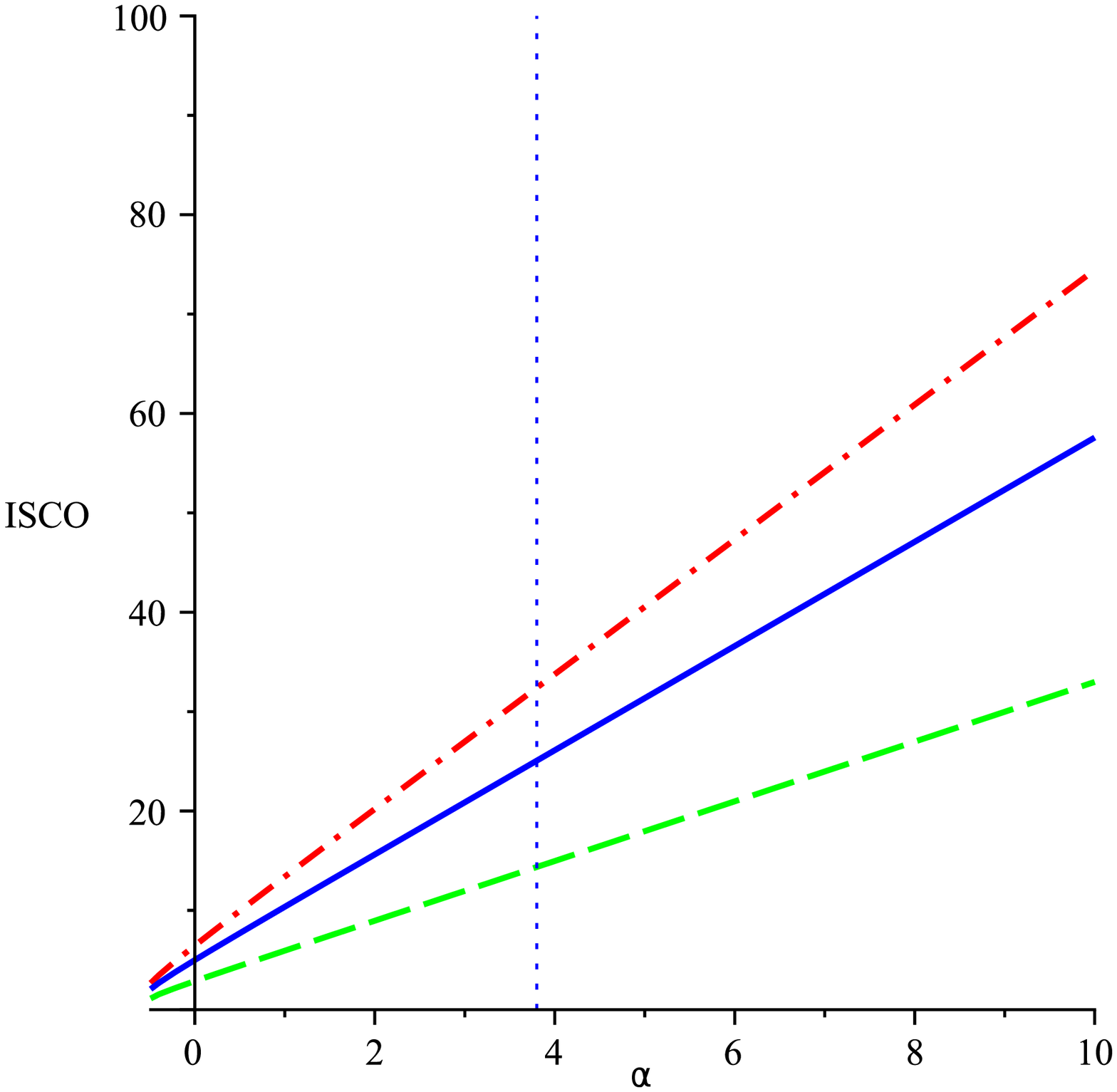}
   \caption{\label{ISCOS}The vertical axis presents the $x$ value at the ISCO. For any chosen value for parameter $\alpha$, the solid line is ISCO for $\beta=0$, the dot-dashed line is ISCO for $\beta_{\rm max}$, and the dashed line is ISCO for $\beta_{\rm min}$. In all cases, ISCO is a monotonically increasing function of $\alpha$. Besides, one can see that the place of ISCO is also a monotonic increasing function of $\beta$. For example, if we look at the plot for any fixed value of $\alpha$, like the vertical dotted line, the intersections of this line with all lines shows the place of ISCO for a fixed $\alpha$, but different values of $\beta$. Therefore, one can see the place of ISCO for the negative value of $\beta$ (green line), is less than $\beta=0$ (blue one), and it is also less than a positive value (red line). Thus, the ISCO is also a monotonically increasing function of $\beta$.}
\end{figure*}

It turns out that the maximum value of $\beta$ is a monotonically decreasing function of $\alpha$. However, the place of ISCO for a maximum of $\beta$ is a monotonically increasing function of $\alpha$ (see Figure \ref{ISCOS}).
In addition, the minimum of $\beta$ is an decreasing function of $\alpha$ from $\alpha \sim -0.5528$ to $\alpha=-0.5$, and from this value it is monotonically increases.  

To summarize, the minimum of $\beta$ is obtained in the case $\alpha=-0.5$, and the maximum of $\beta$ is reached for $\alpha \sim -0.5528.$ In Table \ref{T1r}, the values of minimum and maximum of $\beta$ for various chosen values of $\alpha$ are presented  \footnote{Note that in the third row it is calculated for some $\alpha$ very close to $-0.5$.}. Besides, the places of ISCO in the cases of maximum, minimum, and vanishing external quadrupole distortion parameter $\beta$ are shown. We have seen for $\alpha=\beta=0$, the place of ISCO for Schwarzschild is recovered at $x=5$. We should mention that $\beta$ is a parameter chosen for the entire space-time; however, when $\beta$ is outside of the bounds, there will be no circular orbits at the given $x$ (similar to $r<3M$ in Schwarzschild space-time).
\subsection{Circular orbits for the light-like trajectory}
In this case, $\epsilon=0$, and the effective potential \eqref{Vei} for light-like geodesics is reduced to

\begin{align}\label{Veilight}
    V_{\rm Eff}= \left( \frac{x-1}{x+1} \right)^{(\alpha+1)} e^{2\hat{\psi}} \left[ \frac{L^2e^{2\hat{\psi}}}{M^2(x+1)^2} \left( \frac{x-1}{x+1} \right)^\alpha\right].\,
\end{align}
In fact, a straightforward analysis of the effective potential shows that its first derivative vanishes for

\begin{align}\label{curveplight1}
\alpha=\frac{1}{2}(2\beta x^3-2\beta x+x-2),
\end{align}
where for $\alpha=\beta=0$, it reduces to the Schwarzschild value $x=2$ equivalently to $r=3M$ in the standard Schwarzschild coordinates \eqref{transf1}. Furthermore, the relation \eqref{curveplight1} is the limiting curve for time-like circular geodesics, curve $\mathnormal{l}_1$ \eqref{curve1}, which is written in terms of $\alpha$. By inserting the relation \eqref{curveplight1} for $\alpha$, into the second derivative of the effective potential $(V_{\rm Eff})^{''}$, we obtain a very interesting result. The second derivative vanishes along

\begin{align}\label{qlight}
\beta = -\frac{1}{2(3x^2-1)}.
\end{align}
This expression is also the minimum of the curve \eqref{curveplight1}. Surprisingly, this means that for some negative values of $\beta$ we have a bound photon orbit in the equatorial plane in this space-time, which is not the case nor in Schwarzschild spacetime, neither in $\rm q$-metric. In fact, this arises due to the existence of quadrupole related to the external source.
From this relation \eqref{qlight} one can find the negative values of quadrupole which lead to having ISCO for light-like geodesics

\begin{align}
\beta\in (-\frac{1}{4},0).    
\end{align}
There is no surprise that the minimum value of $\beta$ in the case of light-like trajectories coincides with the minimum value of $\beta$ in the case of time-like trajectories, which occurs for the choice of $\alpha=-0.5$, see Table \ref{T1r}.


\subsection{Revisit circular geodesics in \rm q- metric} \label{Cqmetric}
In this part, following the discussion above to have a comparison, we briefly revisit circular motion on the equatorial plane in the $\rm q$-metric with this slightly different approach from the studies in the literature for example \cite{PhysRevD.93.024024}.


\subsubsection{Time-like geodesics in \rm q- metric}
\begin{figure*}
\centering
\includegraphics[width=0.4\textwidth]{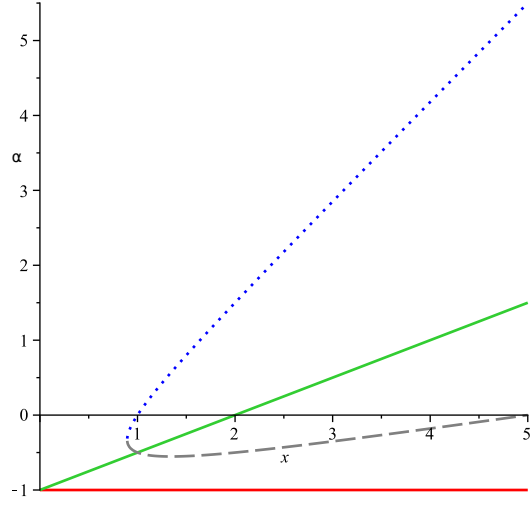}
   \caption{\label{4curveforp}Solid lines $\alpha_{2}$ and $\alpha_{1}$ show the valid region, where dashed line $\alpha_{-}$ always lies in this region and dotted line $\alpha_{+}$ is out of this region for any choice of $\alpha$.}
\end{figure*}

In this case, the specific angular momentum \eqref{angmom} is reduced to

\begin{align}
L^2=\frac{M^2(x+1)^{\alpha+2}(1+\alpha)}{(x-1)^\alpha\left[x-2(1+\alpha)\right]}.\label{angmomi}
\end{align}
An analyzing of $\frac{d}{dx}L^{2}$, like the previous case, shows the vertical asymptote to this function for $x>1$ is


\begin{align}\label{curvep1}
\mathnormal{k}_{1}:=\frac{x}{2}-1.
\end{align}
Also, $L^2$ vanishes along this curve

\begin{align}\label{curvep2}
\mathnormal{k}_{2}:=-1.
\end{align}
This value is the infimum value for $\alpha$, since regarding the domain of $\alpha=(-1,\infty)$, for $\alpha=-1$ the space-time will be flat as mentioned before.
The region between curves $\mathnormal{k}_1$ \eqref{curvep1} and $\mathnormal{k}_2$ \eqref{curvep2}, defines the valid range for $\alpha$, considering $L^2$ is a positive function. A direct calculation for the extrema of $L^2$, leads to


\begin{align}\label{LLp0}
\alpha{\pm}=\frac{3}{4}x-1 \pm \frac{1}{4}\sqrt{5x^2-4}.
\end{align}
Since for the domain of our interest $x>1$, this relation $5x^2-4>0$ satisfies, therefore we always have two curves $\alpha_{-}$ and $\alpha_{+}$. In fact, $\alpha_{+}$ lies out of the valid region between $k_{1}$ and $k_{2}$, for all $x>1$. However, $\alpha_{-}$ intersect with $k_{1}$ and enter to the region at $x=1$, where $\alpha_{-}(1)=-\frac{1}{2}$, and always remain inside this region (see Figure \ref{4curveforp}).

One can show that $\alpha_{-}$ has a minimum inside this region at $x=3\frac{\sqrt{5}}{5}\approx 1.342$, where $\alpha^{\rm min}_{-}=-1+\frac{\sqrt{5}}{5}\approx -0.5528$, and after this point, $\alpha_{-}$ is a monotonically increasing function. So in this case, the domain of $\alpha$ is $[-1+\frac{\sqrt{5}}{5}, \infty).$

\begin{figure*}
\centering
\includegraphics[width=0.35\textwidth]{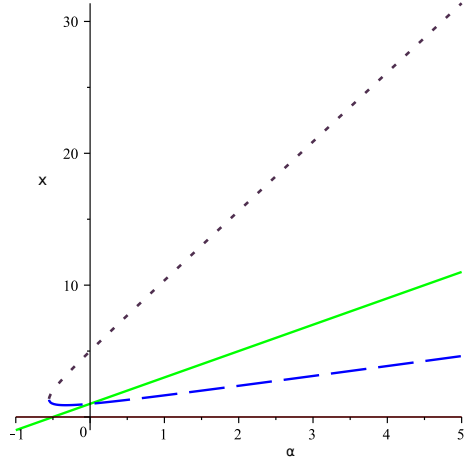}
   \caption{\label{4curveforxqzero}The green line $x_{1}=2\alpha+1$ and $x_{2}=0$, show the valid region, where the dashed line is the place of ISCO $x_{+}$, and dots line $x_{-}$ is outside, for any choice of $\alpha$.}
\end{figure*}

For obtaining ISCOs, we rewrite equation $L^2_{,x}=0$ in terms of $x$,

\begin{align}\label{LLx0}
x_{\pm}=3+3\alpha \pm \sqrt{\Delta}, \quad \quad \Delta=5\alpha^2+10\alpha+4.
\end{align}
By using the above analysis for $\alpha_{\pm}$, from \eqref{curvep1} and \eqref{curvep2} we obtain

\begin{align}
 x_{1}=&2\alpha+1,\\
 x_{2}=&0.\ 
\end{align}
If we plot them with $x_{+}$ and $x_{-}$ \eqref{LLx0} together, we see that the corresponding valid ISCOs are obtained by $x_{+}(=x_{\rm ISCO})$ (see Figure \ref{4curveforxqzero}). Furthermore, this relation \eqref{LLx0} shows that $\Delta\geq 0$ is equivalent to $\alpha \in [-\infty, -1-\frac{\sqrt{5}}{5}]\cup [-1+\frac{\sqrt{5}}{5} , +\infty)$, where only the second part, meaning $[-1+\frac{\sqrt{5}}{5} , \infty)$, lies in the valid region. Consequently, this gives us no new information on the domain of $\alpha$ more than what we have obtained before.

\begin{figure*}
\centering
        \includegraphics[width=0.35\textwidth]{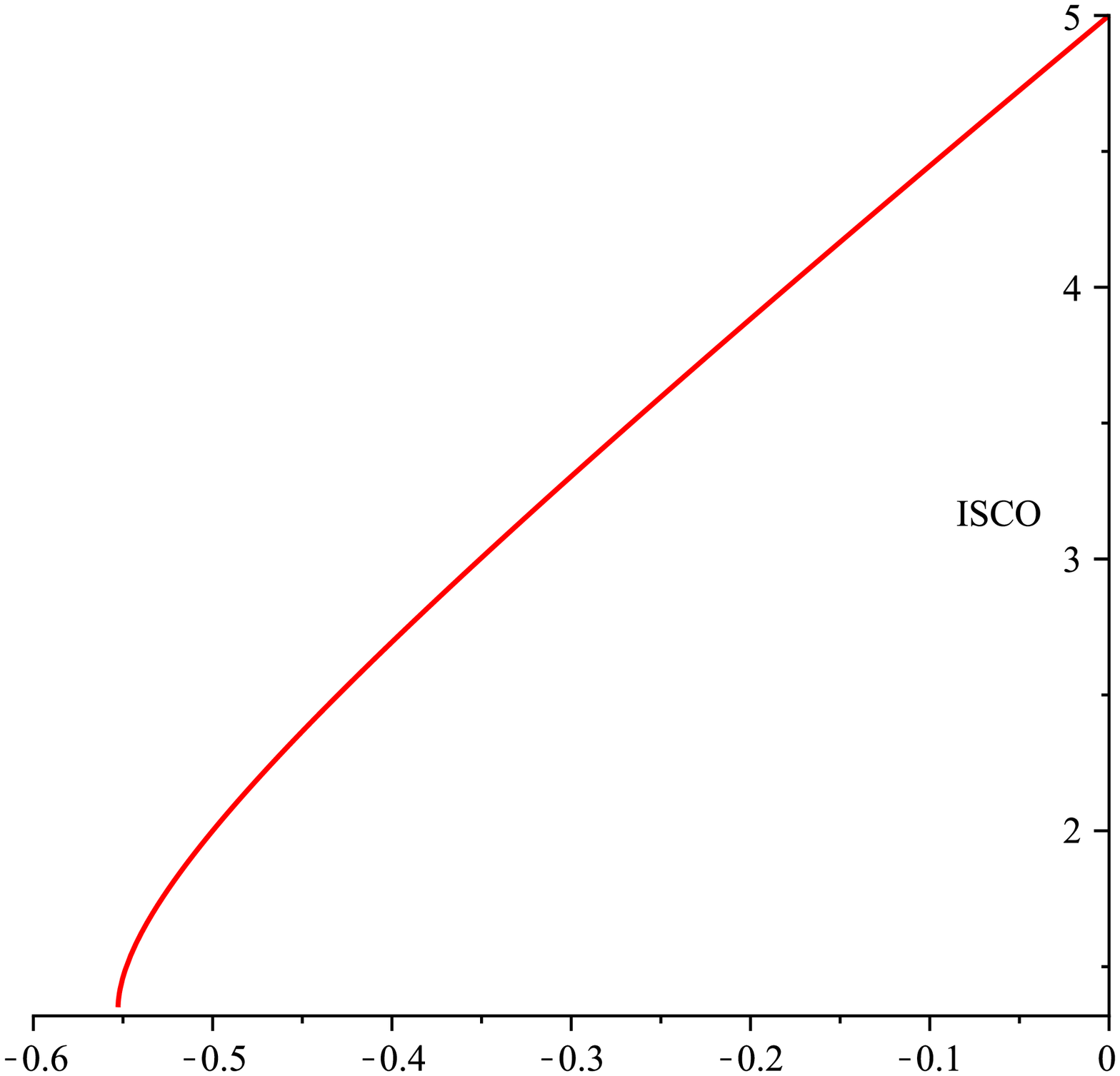}
         \includegraphics[width=0.35\textwidth]{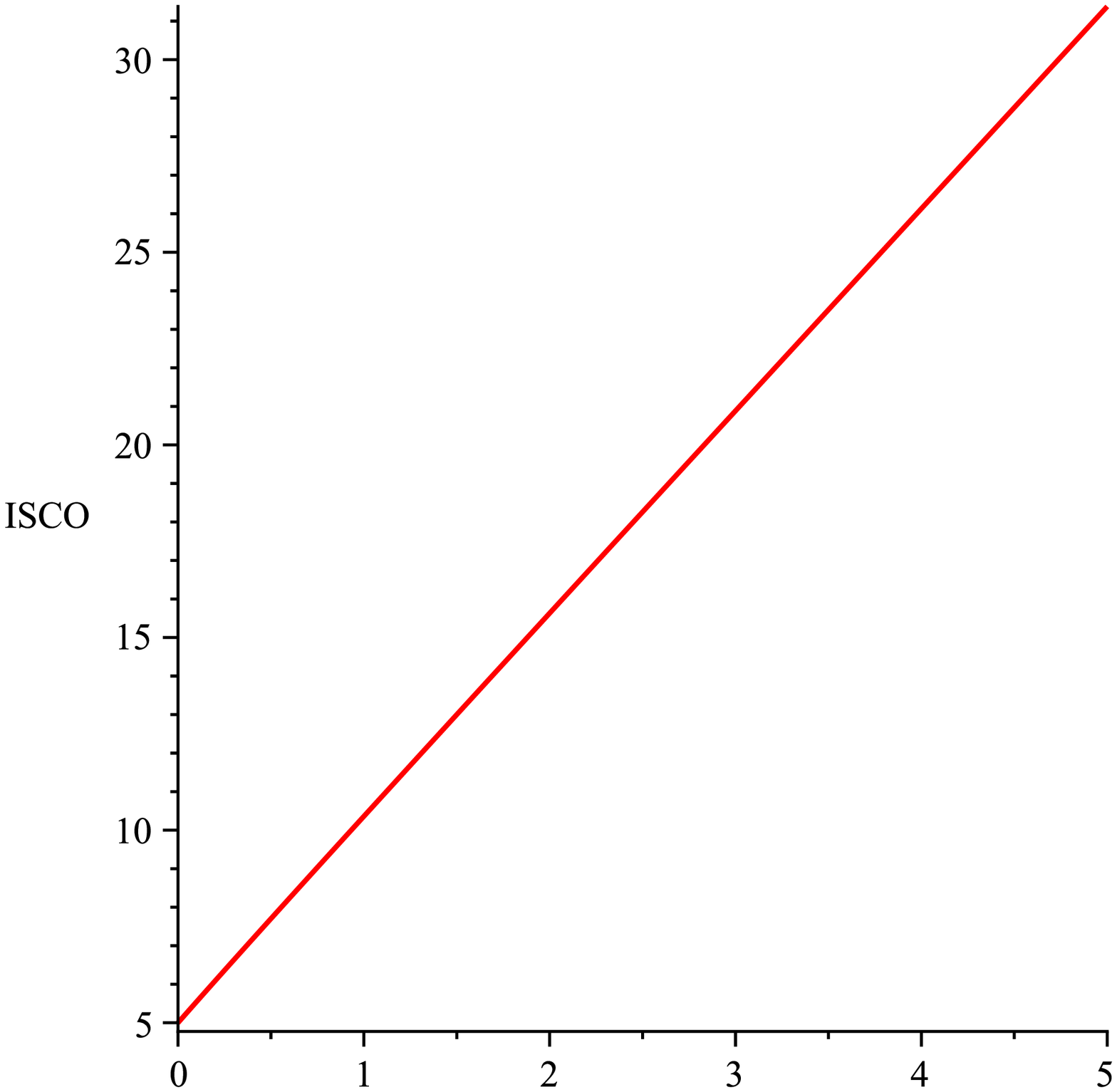}
    \caption{\label{iscopqzero}The vertical axis presents the $x$ value at the ISCO. The plots show the evolution of ISCO with $\alpha$, where the domain of $\alpha$ is $[-0.5528, +\infty)$. The value of ISCO at $\alpha=-0.5528$ is $1.3416$. The place of ISCO for $\alpha<0$ is below the Schwarzschild value $x=5$ and for $\alpha>0$ is upper.}
\end{figure*}
Again, we can see how ISCOs positions evolve as $\alpha$ increases by plotting $x_{+}(=x_{\rm ISCO})$.
the place of ISCO in Schwarzschild is at $x=5$ (equivalent to $r=6M$). For a negative $\alpha$, ISCO is closer to the horizon $x=1$, and for a positive $\alpha$ the place of ISCO is going farther, see Figure \ref{iscopqzero}.

\subsubsection{Light-like geodesics for \rm q-metric}
In this case, the effective potential \eqref{Veilight} is reduced to

\begin{align}\label{Vilight}
    V_{\rm Eff}= \left( \frac{x-1}{x+1} \right)^{(\alpha+1)} \left[ \frac{L^2}{M^2(x+1)^2} \left( \frac{x-1}{x+1} \right)^\alpha\right],\,
\end{align}
and the straightforward calculation shows that its first derivative vanishes along

\begin{align}
x=2+2\alpha.
\end{align}
So for any chosen value of $\alpha$, one obtains a value for $x$. Of course, in the case of $\alpha=0$, Schwarzschild metric, we obtain $x=2$ or equivalently $r=3M$ with the transformation law \eqref{transf1}. Moreover, the sign of the second derivative of the effective potential for this value of $x$, and any chosen value for $\alpha$, unlike the previous case indicating this circular motion is unstable.


\section{Summary and conclusion}\label{discuss}


%

This paper has presented a generalized $\rm q$-metric for the relatively small quadrupole moment via Weyl's procedure. This metric explains the exterior of a deformed body locally in the presence of an external distribution of matter up to the quadrupole. It contains three free parameters: the total mass, deformation parameter $\alpha$, and the distortion parameter $\beta$, referring to the central object's quadrupoles and its surrounding mass distribution, where in the case of the circular motion these two quadrupole parameters are not independent.

We propose that this class has promising features, and it may serve to link the metric to the physical nature of the deformed compact object and its surrounding via its parameters. Besides, this generalization is worth considering to have more vacuum metrics available, for example, to study the interplay of the vacuum solution with an external field. Therefore, it may be of some interest to investigate how the external field affects the geometry and geodesics of the $\rm q$-metric. In fact, due to its relatively simple form, it is possible to use this in (semi-) analytical models of astrophysical relevance, like the analysis of geodesic motion, accretion disc models, quasi-normal modes, quasi-periodic oscillations, and more.






Furthermore, we carried out a characterization of quadrupole parameters' impact via studying the circular geodesics on the equatorial plane, as well as the region of the parameters
for which circular orbits can exist. Some of the examples are listed in Table \ref{T1r}. In consequence, we found out for each choice of $\alpha$ there are ISCOs for time-like geodesics for $\beta \in [\beta^{\alpha}_{\rm min}, \beta^{\alpha}_{\rm max}]$ such that in general $\beta_{\rm min}\approx -2.5\times10^{-1}$ at $\alpha=-0.5$ and $\beta_{\rm max}\approx 4.1\times10^{-3}$ at $\alpha=-0.5528$. Besides, the place of ISCO is closer to the horizon for negative quadrupole moments on the contrary to positive quadrupole moments.

An interesting result is that there is a bound orbit for light-like geodesics on the equatorial plane, which is not the case neither in Schwarzschild nor in $\rm q$-metric. The key point is, this bound orbit's existence directly is reflected into the having negative quadrupole of external matter and provides the range for $\beta\in(-0.25,0)$ in this case. In fact, most of our information about the astrophysical environment is obtained from electromagnetic radiation and consequently by studying the null geodesics; therefore, this result is of great astrophysical interest like studying shadow. 

As a final remark we would like to mention that the obtained results admit a generalization to the case of Stationary $\rm q$-metric in the external static gravitational field. In addition, test particles' motion expected to be chaotic in the equatorial plane for some combinations of parameters $\alpha$ and $\beta$, since a perturbation in the gravitational or the electromagnetic field
generally leads to chaos. This also might be an interesting point to investigate in the future works. 

The next step of this work could be a study on off-equatorial time-like and light-like geodesics. Studying the topological implication of this background from the  mathematical or astrophysical perspective. Furthermore, considering the metric is valid locally, it seems reasonable to use an approach like the perturbative matching calculation to describe the external universe \cite{PhysRevD.69.084007}, which may be the next stage of this study. In addition, study stability of this solution is of some interest. Further, the model can reproduce some of the relevant features of the numerical simulation in the astrophysical setting like MHD simulation of the accretion discs. We expect future work in this field to be guided by astrophysics questions and other areas where strong-field gravitational theory applies.

\begin{table*}
 \caption{The minimum and the maximum of $\beta$ and the place of ISCO for different values of $\alpha$. The letter $\alpha$ appearing as the upper index in $\beta$ and $x$ means that these quantities are calculated at a fixed value of $\alpha$ in the left column.\label{T1r}}
 \begin{tabular}{|p{1.5cm}||p{3cm}|p{2cm}|p{3cm}|p{2cm}||p{2cm}|} 
 \hline 
$\alpha$&$\beta^{\alpha}_{\rm min}$ &$ x^{\alpha}_{\beta-\rm min}$&$\beta^{\alpha}_{\rm max}$&$ x^{\alpha}_{\beta-\rm max}$&$x^{\alpha}_{\beta=0}$\\
 \hline \hline
  -0.5528 &  -0.0006262  & 1.333070   & 0.0040976 &   1.93984 & 1.40333  \\\hline
   -0.526  &   -0.0443754 & 1.15685  &  0.0028270 &  2.30093 & 1.77325  \\\hline
   -0.5   & -0.2499996   &  1.000001 & 0.00217281  &  2.58141 & 2.00000 \\\hline
  -0.49   & -0.1757730   & 1.13203 & 0.0019932 & 2.68029   &  2.07818  \\\hline
   -0.4   & -0.0805014   & 1.55038 & 0.0011090 & 3.47165   &  2.69443  \\ \hline
   0      & -0.0209443  & 2.87940 & 0.0002927 & 6.45602    &  5.00000  \\ \hline
  0.5    & -0.0086651   & 4.42340 & 0.0001202 & 9.95281   &  7.70156  \\ \hline
  1      & -0.0047632   & 5.94338 & 0.0000659 & 13.38972  &  10.35890  \\ \hline
  10     & -0.0001533  & 32.98501& 0.0000021 & 74.44744   &  57.57641  \\ \hline
 \hline
 \end{tabular}
 \end{table*}

\acknowledgments{The author gratefully acknowledges Prof. Hernando Quevedo, Prof. Domenico Giulini, Prof. Claus Laemmerzahl, Dr. Eva Hackmann, and Dr. Audry Trova for valuable discussions, also the support of the Cluster of Excellence EXC-2123 Quantum Frontiers and the research training group GRK 1620 Models of Gravity by DPG.}







\appendix
\section{Christoffel symbols} \label{circulargeo}

The geodesic equation in an arbitrary space-time is described by 

\begin{align}\label{geoequ}
\ddot{x}^{\mu}+\Gamma^{\mu}_{\nu \rho}\dot{x}^{\nu}\dot{x}^{\rho}=0\nonumber,
\end{align}
where "over-dot" notation is used for derivations with respect to the affine parameter, $\dot{x}^{\mu}$ is the four-velocity, and $\Gamma^{\alpha}_{\beta \gamma}$ are the Christoffel symbols, which in this space-time read as follows 


\begin{align}
\Gamma^t_{tx}&=\frac{(1+\alpha)}{x^2-1}+\hat{\psi}_{,x},\nonumber\\
\Gamma^t_{ty}&=\hat{\psi}_{,y},\nonumber\\
\Gamma^x_{tt}&= \frac{e^{4\hat{\psi}-2\hat{\gamma}}}{M^2}\left(\frac{x-1}{x+1}\right)^{2\alpha+1}\left(\frac{x^2-y^2}{x^2-1}\right)^{\alpha(\alpha+2)}\nonumber\\
&\left[\frac{1+\alpha}{(x+1)^2}+\left(\frac{x-1}{x+1}\right)\hat{\psi}_{,x}\right],\nonumber\\
\Gamma^x_{xx}&=-\frac{1+\alpha}{x^2-1}+\frac{\alpha(\alpha+2)(y^2-1)(1+2x)}{(x^2-1)(x^2-y^2)}\nonumber\\
&+\hat{\gamma}_{,x}-\hat{\psi}_{,x},\nonumber\\
\Gamma^x_{xy}&= -\frac{y\alpha(\alpha+2)}{x^2-y^2}+\hat{\gamma}_{,y}-\hat{\psi}_{,y},\nonumber\\
\Gamma^x_{yy}&=\left(\frac{1-x}{1-y^2}\right)\left[1-\frac \alpha{x-1}+(\hat{\gamma}_{,x}-\hat{\psi}_{,x})(x+1)\right]\nonumber\\
&-\frac{x}{(x^2-y^2)}\alpha(\alpha+2),\nonumber\\
\Gamma^x_{\phi\phi}&=e^{-2\hat{\gamma}}(1-y^2)\nonumber\\
&\left(\frac{x^2-y^2}{x^2-1}\right)^{\alpha(\alpha+2)}\left[1+\alpha-x+\hat{\psi}_{,x}(x^2-1)\right],\nonumber\\
\Gamma^y_{tt}&=\frac{e^{4\hat{\psi}-2\hat{\gamma}}}{M^2}\left(\frac{1-y^2}{x^2-1}\right)\left(\frac{x^2-y^2}{x^2-1}\right)^{\alpha(\alpha+2)}\hat{\psi}_{,y},\nonumber \\
 \Gamma^y_{xx}&=\frac{y^2-1}{x^2-1}\left[\frac{y\alpha(\alpha+2)}{x^2-y^2}+\hat{\gamma}_{,y}-\hat{\psi}_{,y}\right], \nonumber\\
\Gamma^y_{xy}&=\frac{x-1-\alpha}{x^2-1}+\left(\frac{1-y^2}{x^2-y^2}\right)x\alpha(2+\alpha)\nonumber\\
&+\hat{\gamma}_{,x}-\hat{\psi}_{,x},\nonumber \\
\Gamma^y_{yy}&=\frac{y}{1-y^2}+\frac{y\alpha(\alpha+2)}{x^2-y^2}+\hat{\gamma}_{,y}-\hat{\psi}_{,y},\nonumber \\
\Gamma^y_{\phi \phi}&=\left(\frac{x^2-y^2}{x^2-1}\right)^{\alpha(2+\alpha)}(1-y^2)e^{-2\hat{\gamma}}\nonumber\\
&\left[y+\hat{\psi}_{,y}(1-y^2)\right],\nonumber \\
\Gamma^\phi_{\phi x}& = \frac{x-1-\alpha}{x^2-1}-
\hat{\psi}_{,x},\nonumber\\
\Gamma^\phi_{\phi y}& = \frac{y}{y^2-1}-\hat{\psi}_{,y}.\nonumber\,
\end{align}

\bibliographystyle{unsrt}
\bibliography{bibb}


%


\end{document}